\PassOptionsToPackage{shortlabels}{enumitem}

\documentclass[preprints,article,accept,moreauthors,pdftex]{Definitions/mdpi}
\firstpage{1} 
\pubvolume{1}
\issuenum{1}
\articlenumber{0}
\pubyear{2022}
\copyrightyear{2022}
\datereceived{} 
\dateaccepted{} 
\datepublished{} 
\hreflink{https://doi.org/} 
\pdfoutput=1

\usepackage[utf8]{inputenc}

\usepackage{url}
\usepackage{xspace} 

\usepackage{color}
\usepackage{booktabs}
\usepackage{longtable,ltcaption,array}
\usepackage{todonotes}
\usepackage{float} \floatstyle{boxed} \restylefloat{figure}
\usepackage{graphicx}
\usepackage{amsmath,mathtools}
\usepackage{amssymb}
\usepackage[olditem,oldenum]{paralist}

\graphicspath{figures}

\renewcommand{\vec}[1]{\mathbf{#1}}

\Title{On the Security of Offloading Post-Processing \\ for Quantum Key Distribution}

\TitleCitation{On the Security of Offloading Post-Processing for Quantum Key Distribution}

\Author{Thomas Lor{\"u}nser $^{1,}$*\orcidA{},
Stephan Krenn $^{1}$\orcidB{},
Christoph Pacher $^{1,2}$ \orcidC{} and
Bernhard Schrenk $^{1}$\orcidD{}}

\AuthorNames{Thomas Lor{\"u}nser,Stephan Krenn,Christoph Pacher and Bernhard Schrenk}

\AuthorCitation{Lor{\"u}nser, T.; Krenn, S.; Pacher, C.; Schrenk, B.}

\address{%
$^{1}$ \quad AIT Austrian Institute of Technology, Giefinggasse 4, 1210 Vienna, Austria\\
$^{2}$ \quad fragmentiX Storage Solutions GmbH, Plöcking 1, 3400 Klosterneuburg, Austria}

\corres{Correspondence: Thomas.Loruenser@ait.ac.at; Tel.: +43-664-8157857}

\abstract{%
Quantum key distribution (QKD) has been researched for almost four decades and is currently making its way to commercial applications.
However, deployment of the technology at scale is challenging, because of the very particular nature of QKD and its physical limitations.
Among others, QKD is computationally intensive in the post-processing phase and devices are therefore complex and power hungry, which leads to problems in certain application scenarios.
In this work we study the possibility to offload computationally intensive parts in the QKD post-processing stack in a secure way to untrusted hardware.
We show how error correction can be securely offloaded for discrete-variable QKD to a single untrusted server and that the same method cannot be used for long distance continuous-variable QKD.
Furthermore, we analyze possibilities for multi-server protocols to be used for error correction and privacy amplification.
Even in cases where it is not possible to offload to an external server, being able to delegate computation to untrusted hardware components on the device could improve the cost and certification effort for device manufacturers.
}


\begin{document}

\section{Introduction%
  \label{sec:introduction}%
}

Quantum key distribution (QKD) was invented almost 40 years ago 
and is currently a more vital field of research than ever.
With commercial impact on the horizon, application of QKD is gaining substantial momentum and the technology is expected to be deployed on large scale in the upcoming years.
This is true for both, terrestrial applications as well as to space.

QKD is the only known information-theoretically secure primitive for key exchange and can be considered as part of the quantum-safe toolbox to build long-term secure ICT systems which even resist quantum computer threats.
However, its wide adoption is still hampered by various challenges which have to be overcome to make QKD practically relevant and facilitate commercial adoption.
On the one hand, research is thus continuously improving protocols and optics/electronics to achieve better bandwidth and distance, as well as co-existence with existing infrastructure.
On the other hand, miniaturization and electro-optical integration are important topics to make the technology more reliable and cost-effective.

Complementary to these efforts, our work focuses on the possibility to offload (outsource) computationally intensive tasks in the QKD post-processing phase to external infrastructure without compromising the overall security.
Being able to outsource these tasks to external data centers allows for simpler and less power-hungry devices in the field, resulting in more versatile applications for QKD. 

\textbf{Related work.}
Improving the efficiency and throughput of the post-processing phase is still an interesting challenge in the context of QKD.
Scientific and industrial research and development initiatives are focusing on algorithmic improvements to reduce computational effort (c.f. \cite{Martinez-Mateo2015,Pedersen2015,e23111440,PAL+2016}), on extending the local computational resources with specialized hardware for high-performance computing and or graphics processing units \cite{MPH+2013,Wang2018a,Li2020}, and on developing dedicated hardware designs in field programmable hardware designs as co-processing units and local deployments \cite{Yang20,9376906}.

\textbf{Contributions.}
In this work, we contribute to these efforts via a complementary approach, by presenting novel methods for offloading (or outsourcing) the most expensive parts of the QKD post-processing stack.
To do so we combine our expertise from QKD and cryptography in order to motivate the problem and show the benefits as well as present protocols and barriers.

More precisely, we present and analyze protocols for outsourcing information reconciliation to external and untrusted environments, therefore facilitating new application scenarios, e.g., usage in low-power access networks. We furthermore discuss possibilities to outsource the privacy amplification step, which could further help to reduce the required processing power in QKD nodes.
Additionally, we present use cases in order to undermine the practical relevance of the novel developed methods.

\textbf{Outline of the work.}
In Section \ref{sec:qkd} we present and discuss quantum key distribution and the required steps for post-processing, as well as the motivation for offloading computationally intense tasks.
In Section \ref{sec:reconciliation} we review information reconciliation in detail and present new protocols which allow for outsourcing them in a secure way as well as an impossibility result.
In Section \ref{sec:privacyamp} we are analyzing the potential of outsourcing privacy amplification and propose new methods towards this directions.
Potential use cases for application of the proposed solution are then discussed in Section \ref{sec:usecases} and the concluding remarks are given in Section \ref{sec:conclusions}.

\section{Quantum key distribution}\label{sec:qkd}

Contrary to most other cryptographic primitives, QKD is a cryptographic key-agreement protocol which derives its security from the physical layer, i.e., it uses a quantum channel to exchange quantum information which cannot be perfectly copied or eavesdropped according to the laws of quantum mechanics.
In prepare and measure QKD protocols, so called quantum bits (qubits) are encoded and transmitted over a quantum channel.
Typically, the qubits are encoded on photons and the transmission channels are either fiber optics or free space.
Finally, the qubits are measured at the receiver and decoded.
From the measurement of quantum bits classical information is derived and all following steps are done in the classical domain.
However, due to their interaction with the environment and/or eavesdroppers photons are subject to perturbation and absorption.
To detect and cope with these modifications in the transmission channel, post-processing steps have to be applied in order to get the full key agreement primitive with practical correctness and secrecy guarantees.

The outstanding property of QKD is that it is an information-theoretic secure (ITS) and universally composable (UC) key agreement protocol \cite{Muller_Quade_2009}, given that its  classical  communication  is performed over an authentic channel (note that all key-agreement protocols are insecure over non-authentic channels). ITS message authentication codes based on universal hashing \cite{WegmanCarter} which in the first round use preshared keys and in later rounds QKD keys from previous rounds are a means to generate an ITS authentic \cite{Portmann} channel. 
Thus, QKD is a very powerful cryptographic primitive which cannot be realized with non-quantum protocols.

\subsection{QKD post-processing}
\label{ssec:post-processing}

QKD comprises two phases to arrive at a key agreement with strong correctness and secrecy guarantees. 
First, qubits are randomly generated on one side, transmitted over the optical quantum channel, and measured on the other side to generate the so called raw key.
In the second phase, a non-quantum (classical) post-processing protocol is executed to agree on identical keys (correctness) on both ends of the transmission line, and to render useless any information a potential attacker could have learned by attacking the transmission phase (secrecy).

In detail, the steps to extract a secure key from the raw data of the transmitted quantum bits are as follows:
\begin{enumerate}[(i)]
  \item 
	\textbf{Sifting} removes non-relevant information from the raw key; e.g., in conjugate coding protocols, events prepared and measured in different bases are deleted.
Also, events not received by Bob are discarded in discrete-variable protocols (cf. Section~\ref{sec:reconciliation}).
  \item 
	\textbf{Error estimation} determines an upper bound on the information leaked to an adversary on the quantum channel and can provide information to optimize the subsequent information reconciliation.
Although more advanced methods have been proposed in the literature, this is typically done by cut-and-choose methods.
Additionally, the idea of using a confirmation phase to replace error estimation was proposed by
L{\"u}tkenhaus~\cite{lutken99}.
  \item 
	\textbf{Information reconciliation}---which often uses methods from forward error correction---aims at correcting all errors in the remaining raw key so that sender and receiver should obtain identical keys. 
The classical (non-quantum) messages exchanged in this process must not leak information on the final key.
Typically, the leakage is tracked and treated during the privacy amplification step.
  \item 
	\textbf{Confirmation} detects non-identical keys (for which information reconciliation has failed) with probability close to one.
If non-identical keys are detected, the parties either go back to the information reconciliation step or abort the QKD protocol. 
  \item 
Finally, \textbf{privacy amplification} eliminates the information leaked during all protocol steps (quantum and classical) from the final key by running a (strong) randomness extraction protocol between the peers.
\end{enumerate}
All processing steps together enable Alice and Bob to agree on a final key which is $\epsilon$-close to an ideal key.

Various optimizations of the above key agreement process have been proposed in the past, either for efficiency reasons or implementation aspects, but this generic structure is typically followed in one way or the other.

\subsection{Motivation to offload post-processing}

From a computational perspective, information reconciliation is by far the most computationally intense task in the stack and can be limiting the throughput in high-speed systems \cite{Yuan18,Ren2021}. 
The second most computationally demanding task is privacy amplification \cite{Wang2018a}.
The rest of the protocol steps are rather simple tasks and can be executed in real-time even on embedded platforms.

Therefore, we introduce and study the idea of offloading these tasks from devices by outsourcing computation to untrusted or less trusted hardware in an ITS secure way.
The ability to outsource information reconciliation (IR) and potentially privacy amplification (PA) would enable new applications scenarios for both, the access network and the transmission systems.
On top of these, satellite-based QKD can become more versatile, if processing resources can be shifted around more easily. 

The two main advantages gained by offloading processing to external hardware are increased efficiency and flexibility in the use of compute resources---also resulting in a higher energy efficiency---and a reduced attack surface by limiting the number of components dealing with secure key material.

QKD systems are deployed for long-term security and produce large capital expenditure (CAPEX) spending, i.e., they are used over a long period of time.
Putting all the processing power into the devices at build time hinders later updates and prevents the operator to benefit from Moore's law. If the hardware is outsourced, it could be updated during the lifetime of the system with new technologies resulting in further optimized efficiency.
Furthermore, if the hardware need not be trustworthy and certified, cheaper commercial off-the-shelf (COTS) hardware could be used.
It would even be possible to completely outsource it to public cloud infrastructures in the extreme case.

Additionally, time sharing allows for further improvements for certain use cases.
If QKD is used in hybrid encryption protocols to establish session keys \cite{NPW+2008}, high key rates are not needed and sharing the computational resources between links can further reduce CAPEX and also operational cost (OPEX).
Hence, putting the computational expensive tasks into efficient data centers which do not even need to be trusted is very desirable.
It allows for hardware updates and joint management of all QKD workloads in the field with its continuous upgrading probabilities, which is especially favorable for operators of QKD networks. 
Because the communication overhead is minimal compared to the computational one, a clear advantage in terms of energy efficiency arises and the gained flexibility in managing tasks is very advantageous.

Furthermore, the proposed approach could also be used within a system.
Treating parts of the system as untrusted could eventually provide the possibility for system updates without compromising system certification and help to reduce OPEX cost during system lifetime.

\section{Outsourcing Information Reconciliation}\label{sec:reconciliation}


As mentioned in Section \ref{sec:qkd}, information reconciliation (IR) is the most demanding task in post-processing of QKD, independent of the protocols being used on the quantum level.
Error correction is computationally intense, because of high error rates encountered in combination 
with constraints on the amount of information disclosed during error correction.
The information revealed during the public discussion must be kept as short as possible to maximize overall system performance, ideally IR works close to the Shannon limit.
If keys have to be processed in real-time, error correction is the bottleneck of post-processing and can introduce substantial problems in resource constraint environments.

On a quantum level, QKD protocols can be divided into two classes--\emph{discrete-variable} (DV) and \emph{continuous-variable} (CV) QKD--which also result in different requirements on information reconciliation.
In DV-QKD protocols, e.g., BB84~\cite{BB84}, qubits are measured by single photon detectors.
Due to channel attenuation and non-perfect detectors the rate of detected photons is typically orders of magnitudes lower than the rate of prepared photons.
Consequently, in DV-QKD, IR schemes must typically provide the possibility to operate on raw key rates in the order of Kilobit per second~\cite{Grunenfelder2020} up to Megabit per second~\cite{Yuan18}. 
In CV-QKD systems signals are only perturbed but not lost through channel effects, resulting in very high raw key rates but also high error rates compared to discrete variable system.

Additionally, two basic types of IR protocols can be distinguished in QKD systems.
On the one hand, interactive, two-way protocols have been developed for highly efficient correction capabilities near the Shannon limit, with CASCADE~\cite{Brassard1993,Martinez-Mateo2015,Pacher2015} being the most prominent representative. They can achieve smaller leakage than any one-way protocol, however, their practical performance is limited due to their interactive nature by the latency of the classical channel.
On the other hand, forward error correcting schemes have been adopted and developed further to be used in operational regimes encountered in QKD \cite{5503195}.
One-way IR based on low-density parity-check codes (LDPC) is currently the most efficient representative in this category and used in many prototype systems \cite{PhysRevA103062419}.

One-way schemes have many desirable properties when it comes to realization and can easily be parallelized to increase performance.


\subsection{Linear One-way Information Reconciliation}

Before presenting our scheme for offloading, we first explain one-way IR in the context of DV-QKD in more detail and informally define the concept of secure outsourcing for IR.
Traditional error correcting block codes consist of sets of codewords that contain redundant information.
Before sending data over a noisy channel, the data is encoded into codewords.
The contained redundancy can then be used by the receiver to correct the introduced errors.

One-way reconciliation--aka source coding (or compression) with side information--has been studied since the 1970s \cite{slepian-wolf-73,DBLP:journals/tit/WynerZ76}. 
While related to error correcting codes, the idea here is that the data is transmitted over a noisy channel without adding any redundancy. 
Rather, the source additionally sends a compressed version of the data over a \emph{noisefree} channel. The receiver uses the compressed data together with the noisy data (side information) to decode the original data.

More concretely, after sifting Bob has obtained a noisy version of Alice's sifted key, i.e. $\vec{k}_B=\vec{k}_A + \vec{e}$, where $\vec{e}$ denotes the error vector.
Alice and Bob then use a linear block code with parity check matrix $\vec{H}$. 
Alice compresses her sifted key $\vec{k}_A$ with the help of $\vec{H}$ by computing the corresponding syndrome
$$
\vec{s}_A := \vec{k}_A\vec{H}^\top.
$$
Alice sends $\vec{s}_A$ over the noisefree classical channel to Bob.
Bob corrects his sifted key by (approximately) solving the problem of finding a vector $\vec{\hat k}_A$ which among all vectors with syndrome $\vec{s}_A$
 has the smallest Hamming distance to $\vec{k}_B$.

Searching for this vector is computationally hard and is equivalent to solving the standard syndrome decoding problem, which, given $\vec{H}$ and a vector $\vec{s}$ requires to find a vector $\vec{e}$ of minimal weight satisfying $\vec{e}\vec{H}^\top=\vec{s}$.
The equivalence can easily be seen considering that in our case the syndrome decoder is employed for $\vec{e}\vec{H}^\top=\vec{k}_B\vec{H}^\top-\vec{k}_A\vec{H}^\top=\vec{k}_B\vec{H}^\top - \vec{s}_A$.

%
%

\subsection{Protocol for offloading direct reconciliation}

In the context of QKD, if Alice who sends the qubits sends also the syndrome to Bob and Bob corrects erroneous bits to obtain the sifted key of Alice, the protocol is called \emph{direct reconciliation} (DR).
In DV-QKD which is considered symmetric \cite{Scarani2009} the roles of Alice and Bob can also be interchanged during IR, resulting in so-called \emph{reverse reconciliation} (RR).
However, the same is not true for CV-QKD where the direction of the IR protocol does matter for higher transmission rates, as discussed later.

We present a simple scheme for remote (outsourced) information reconciliation called \texttt{REM-IR} (remote IR), which enables the computationally intense step of syndrome decoding to be outsourced to an untrusted party in a secure way.
The idea is to give the error syndrome, i.e., $\vec{s_e}=\vec{s}_B-\vec{s}_A=\vec{k}_B\vec{H}^\top - \vec{k}_A\vec{H}^\top$ to an external party, which returns the error vector $\vec{e}$ with minimal weight satisfying $\vec{s_e}=\vec{e}\vec{H}^\top$.
    
Informally, the protocol is secure because the information leaked by publishing $\vec{s_e}$, $\vec{e}$ and thus $\vec{s_A}$ does not increase the information of Eve about the agreed key string.
The intuition behind is that just learning a bit flip vector of unknown key does not increase the information about the key.
The described protocol is also equivalent to interactive error decoding as introduced in CASCADE \cite{Brassard1993}, which also leaks parity information and error bit locations during the public discussion.

A detailed description of the protocol is shown in Figure \ref{fig:rem-irc} and the security of the protocols is proved in the following. 
\begin{figure}[t]
    \textbf{Protocol \texttt{REM-IR}:}
	    \begin{enumerate}
        \item Alice generates her syndrome as $\vec{s}_A=\vec{k}_A\vec{H}^\top$ and sends it to Bob.
        \item Bob generates his syndrome as $\vec{s}_B=\vec{k}_B\vec{H}^\top$ and calculates the error syndrome as $\vec{s}_e=\vec{s}_B-\vec{s}_A$ 
        \item Bob sends the error syndrome $\vec{s_e}$ to the third party
        \item The third party calculates the error vector $\vec{e}$ corresponding to $\vec{s_e}$, i.e., it searches for the $\vec{e}$ with the minimum weight fulfilling $\vec{s_e}=\vec{e}\vec{H}^\top$ and returns $\vec{e}$ to Bob
	   \item Optional: Bob verifies that $\vec{s_e}=\vec{e}\vec{H}^\top$ and that $\vec{e}$ has low weight, cf. Section~\ref{s:verifiability}
        \item Bob calculates $\vec{\hat{k}}_A = \vec{k_B}+\vec{e}$
    \end{enumerate}
    \caption{\texttt{REM-IR} Protocol}
    \label{fig:rem-irc}
\end{figure}

\begin{Theorem}[Security of \texttt{REM-IR}]
    \texttt{REM-IR} is a secure scheme for offloading direct reconciliation for DV-QKD and does not leak any additional information about the agreed key by public discussion compared to a local IR, i.e., the mutual information between Eve's information and the agreed key is the same as with local IR. 
\end{Theorem}
\begin{proof}
  Let $\vec{K}_A, \vec{K}_B$ be $n$ bit random variables representing correlated sifted keys at Alice and Bob, which are used as input to information reconciliation.
  $\vec{S}_A$ is the random variable representing the syndrome computed by Alice and $\vec{E}$ the random variable for the error introduced on $n$ channel usages.
  The quantum channel between Alice and Bob is then modeled as a binary symmetric channel $BSC(e)$ with quantum bit error probability $e$.
  Let further $L_E^{IR}(\vec{K} | Q)$ be the additional information leaked to Eve about the agreed key $\vec{K}$ during the information reconciliation phase, beyond what Eve already gained during the previous steps of the key exchange.

Moreover, $H(\vec{K}_A) = H(\vec{K}_B) = 1$ for uniformly random input encoding, and mutual information $I_{AB}\coloneqq I(\vec{K}_A;\vec{K}_B) = n(1-H_b(e))$ is defined by the error probability on the channel.
  Thus, the amount of information required to be exchanged during public discussion is $|Q| \ge H(\vec{K}_A | \vec{K}_B) = n H_b(e)$,
	where we assume an ideal reconciliation algorithm which works at the Shannon limit, i.e, equality holds.
	
	Without loss of generality, we assume that Bob will correct his errors and the agreed key will be $\vec{k} = \vec{k}_A$.
  Note here that in DV-QKD type protocols we have $I_{AE} = I_{BE}$ \cite{Scarani2009}, which makes them suitable for direct reconciliation.

  Information leaked and gained by Eve during the protocol is $L_E^{\texttt{REM-IR}}(\vec{K} | \vec{S}_A) = |\vec{S}_A| = n H_b(e)$, equal to the information leaked in local IR.
  Namely, even by revealing $\vec{s_e}$ and therefore $\vec{s}_B$, the information Eve gains about the final key compared to local IR does not increase:
$$
  L_E^{\texttt{REM-IR}}(\vec{K} | \vec{S}_A, \vec{S}_e) = L_E^{\texttt{REM-IR}}(\vec{K} | \vec{K}\vec{H}^\top, \vec{E} \vec{H}^\top) = L_E^{\texttt{REM-IR}}(\vec{K} | \vec{K}\vec{H}^\top) = L_E^{\texttt{REM-IR}}(\vec{K} | \vec{S}_A)\,,
$$
	because additional information gained is only about $\vec{e}\vec{H}^\top$ which is not related to the final string $\vec{k}$.
\end{proof}

Variants of \texttt{REM-IR} could be, e.g., to let Alice and Bob directly send $\vec{s}_A$ and $\vec{s}_B$, respectively, to the third party, who then computes $\vec{s_e} = \vec{s}_B - \vec{s}_A$. This version is equivalent, as also in \texttt{REM-IR} the third party knows all syndromes, i.e., it can compute $\vec{s}_B = \vec{s_e} + \vec{s}_A$ from the publicly known $\vec{s_e}$, $\vec{s}_A$.
Furthermore, to increase the reliability and availability of the results, the computation can be delegated and distributed to an arbitrary number of third parties. The security is not jeopardized by any extended protocol involving more external untrusted parties and serves as a general baseline for such scenarios.

\subsection{On outsourcing reverse reconciliation}

For continuous-variable QKD (CV-QKD) we have different requirements than for discrete-variable QKD which not only impact the modulation schemes but also the information reconciliation.
On the Qbit level CV-QKD uses homodyne detection which allows for soft or hard decoding.
For simplicity we will look only at discrete modulated CV-QKD, in particular binary modulation.
Therefore, in the following we treat the CV-QKD system as hard-input--hard-output channel which operates on classical bit strings.

The idea of reverse reconciliation was introduced by Maurer~\cite{Maurer93} for classical communication and later applied to CV-QKD to overcome the 3dB loss limit~\cite{Grosshans2002}.
In essence, reverse reconciliation is based on one-way error correction in reverse configuration with Bob sending the syndrome $\vec{s}_B$ to Alice, and Alice correcting her bits.

The underlying model is based on two channels, one connecting Alice and Bob and the other connecting Alice and Eve.
Interestingly, if reverse reconciliation is applied in this scenario, a key can still be distilled even if the channel from Alice to Eve is superior to the one from Alice to Bob.
The secret capacity of the channel for reverse reconciliation in \cite{Maurer93} was derived as $C_s = H_b(e+d-2ed) - H_b(e)$, when Alice and Bob have access to a broadcast channel for public discussion.
The bit error probabilities are $e, d$ for the channel from Alice to Bob, and Alice to Eve respectively, and $e+d-2ed$ for the conceptual channel from Bob to Eve.
$H_b$ is the binary Entropy function.

In the classical model of \cite{Maurer93} Shannon Entropy is used in the analysis.
For the case of CV-QKD, the mutual information between Bob and Eve has to be replaced by the Holevo information and finite key effects have to be considered \cite{Leverrier2010}.
However, both refinements do not affect our treatment based on generic BSC channels.

\medskip

For the secret channel capacity argument to be valid, Alice' key have to be kept private, thus, preventing Eve to correct error bits with her key.
With this additional requirement, outsourcing information reconciliation directly as done in \texttt{REM-IR} is not possible.
If both, $\vec{s_e}$ and $\vec{s}_B$ are leaked, $\vec{s}_A = \vec{s_e} + \vec{s}_B$ can be easily computed and the advantage over the conceptual channel is lost, because Eve can correct all errors in the string with Alice and remove uncertainty $H(\vec{K}_E|\vec{K}_B)$.

More formally, the following result shows that \textit{fully offloading} error corrections---i.e., letting a third party perform the entire error correction and simply return $\vec{e}$---to an untrusted party cannot be achieved for both, classical reverse reconciliation and in the quantum setting.

\begin{Theorem}[Impossibility of external syndrome decoding for classical RR]
    For reverse reconciliation in the classical (non-quantum) setting fully offloading syndrome decoding is not possible with positive key rate.
\end{Theorem}
\begin{proof}
  Alice is connected to Bob and Eve over binary symmetric channels (BSC) with error rates $e$ and $d$, respectively.
	She sends out the very same signal $\vec{k}_A$, which is received as $\vec{k}_B$ and $\vec{k}_E$.
  In the case of RR we further have that Alice sends the signal $\vec{k}_A$, but corrects her key for the error received by Bob, i.e., $\vec{k}_B$ is final key $\vec{k}$.

  For binary random input encoding it holds that $H(K_A) = H(K_B) = 1$, and the mutual information $I_{AB}\coloneqq I(K_A; K_B) = 1 - H_b(e)$ is defined by the error probability on the channel.
  $K_A$, $K_B$ and $K_E$ are the binary correlated random variables at Alice, Bob and Eve. 
  The amount of information required to be exchanged during public discussion for reverse reconciliation per channel use is $|Q| \ge H(K_B | K_A) = H_b(e)$. 
  For the proof we assume that optimal codes reaching the Shannon limit are used, i.e., equality holds for syndromes communicated. 
  
  Thus for offloading, any external party taking over the syndrome decoding for $n$ bit keys based on a public $\vec{H}$ needs $|\vec{s_e}| = n H_b(e)$ amount of information to correct for the errors on the AB channel.
  Note here that $\vec{s_e}$ itself does not carry any information about the key, yet still fully defines the error $\vec{e}$.

  We now prove the impossibility in two steps. In a first step 
	  \begin{inparaenum}[(i)]
		  \item we calculate the change in mutual information by offloading the computation of $\vec{e}$ by Alice, and therefore publishing the error syndrome $\vec{s_e}$. 
			In \item we then discuss the influence of discussion needed between Alice and Bob to compute the error syndrome $\vec{s_e} = \vec{s}_A - \vec{s}_B = \vec{k}_A \vec{H}^\top - \vec{k}_B \vec{H}^\top$ in the first place, which clearly needs contributions from both peers.
		\end{inparaenum}

  Furthermore, we know that Eve is not allowed to learn enough information about $\vec{k}$ to correct all errors through its conceptual channel, i.e., $I_{AB}-I_{EB}$ have to be preserved or at least be larger than $0$ to leave Alice and Bob with a secure key.

  In the beginning of the protocol we have $I_{AB} = 1 - H_b(e)$ and $I_{EB} = 1 - H_b(e+d-2ed)$.
  After publishing $\vec{s_e}$ and computing $\vec{e}$ in step $(i)$, the mutual information per bit changes to 
  $I_{AB}^{(i)} = 1$ and $I_{EB}^{(i)} = 1 - H_b(e+d-2ed) + H_b(e) = I_{EA}^{(i)}$.
  This means that with knowledge of $\vec{s_e}$ and implicitly $\vec{e}$, Alice can correct for all errors with Bob but Eve is left with some remaining uncertainty.

  Now, to compute $\vec{s_e} = \vec{s}_A-\vec{s}_B$, another $n H_b(e)$ bits have to be communicated in advance between Alice and Bob (ii), which further impacts the knowledge of Eve about the keys.
  However, after exchanging another $n H_b(e)$ bits about $\vec{k}_B$ in public to compute the error syndrome we still have $I_{AB}^{(ii)} = 1$ but $I_{EA}^{(ii)}$ is also increased to $1$ because
  $I_{EB}^{(i)} + H_b(e) = 1 - H_b(e+d-2ed) + 2H_b(e) > 1$.
  Alice already corrected all errors in step (i), the additional information does not further increase their knowledge.
  Contrary, for Eve the information in step (ii) is useful and further increases the mutual information with Bob up to the maximum of $1$, which means Eve has full knowledge about the agreed key.
  This is due to the fact, that the published information is about independent random variables ${K}_E$ and ${K}_B$ both contributing to the key agreement individually $\vec{k} = \vec{k}_B = \vec{k}_A + \vec{e}$.
  In summary, Eve either learns the key or if step (ii) is encrypted it leads to a negative key balance for QKD in the region of interest with $d \le e$.
\end{proof}

\begin{Corollary}[Impossibility for quantum RR]
  For reverse reconciliation in the quantum setting fully offloading syndrome decoding is not possible with positive key rate.
\end{Corollary}
\begin{proof}
  The quantum case is based on the same assumptions as the classical case and derives by looking at the entropies.
  Bob and Eve are connected to Alice over a quantum channel respectively, whereby $I(\vec{K}_A;\vec{K}_E) \geq I(\vec{K}_A;\vec{K}_B)$ holds.
  We also assume a symmetric system with $H({K}_A) = H({K}_B) = 1$, and consequently $H(\vec{K}_A|\vec{K}_B) = H(\vec{K}_B|\vec{K}_A)$ as well as $H(\vec{K}_A|\vec{K}_E) = H(\vec{K}_E|\vec{K}_A)$, due to Bayes' theorem.
  We also know from the definition of mutual information, that Eve has less uncertainty about the final key, i.e. Bob's key, than Alice $H(\vec{K}_A|\vec{K}_E) \leq H(\vec{K}_A|\vec{K}_B)$. 
  Additionally, because the entropy function is concave we also know that $H(\vec{K}_A|\vec{K}_E) \leq H(\vec{K}_B|\vec{K}_E) \leq H(\vec{K}_A|\vec{K}_E) + H(\vec{K}_A|\vec{K}_B)$.
  Due to Slepian-Wolf's theorem~\cite{slepian-wolf-73} we require Bob to communicate $H(\vec{K}_B|\vec{K}_A)$ bits (e.g. $\vec{s}_B$) to enable Alice to compute the error syndrome.
  Furthermore, we require Alice to eventually publish $H(\vec{K}_A|\vec{K}_B)$ bits (e.g. $\vec{s}_e$) in order to fully outsource error correction also assuming an optimal code.
  Contrary to forward reconciliation, both strings published are useful for Eve, because the information about the error is independent from the bits revealed about Bob's key.

  Thus, with access to this public information, Eve is now able to reduce its uncertainty $H(\vec{K}_B|\vec{K}_E)$ about the key, because 
	\[
	H(\vec{K}_B|\vec{K}_E) \leq H(\vec{K}_A|\vec{K}_E) + H(\vec{K}_A|\vec{K}_B) < 2H(\vec{K}_A|\vec{K}_B)\,,
	\]
	leading to $I(\vec{K}_A;\vec{K}_E) = 1$.
  In essence, after seeing $\vec{s}_e$ and $\vec{s}_B$, Eve can calculate $\vec{s}_A = \vec{s}_B - \vec{s}_e$ and remove all uncertainty $H(\vec{K}_E|\vec{K}_A) < H(\vec{K}_A|\vec{K}_B)$ about $\vec{k}_A$ and subsequently the final key $\vec{k}=\vec{k}_B$.
\end{proof}

However, even with this results in mind, it is unclear if weaker notions of offloading would enable certain levels of partial or assisted secure outsourcing with positive key rates.
An impossibility result for \emph{partial offloading} is hard to formalize, as in an edge case no meaningful computation would be delegated to the untrusted server and the entire error reconciliation would be performed as local operations.
In the following we argue that no obvious or natural approaches for reasonable (partial) delegation of computations do exist.

\begin{figure}[h]
  \centering
  \includegraphics[width=0.9\textwidth]{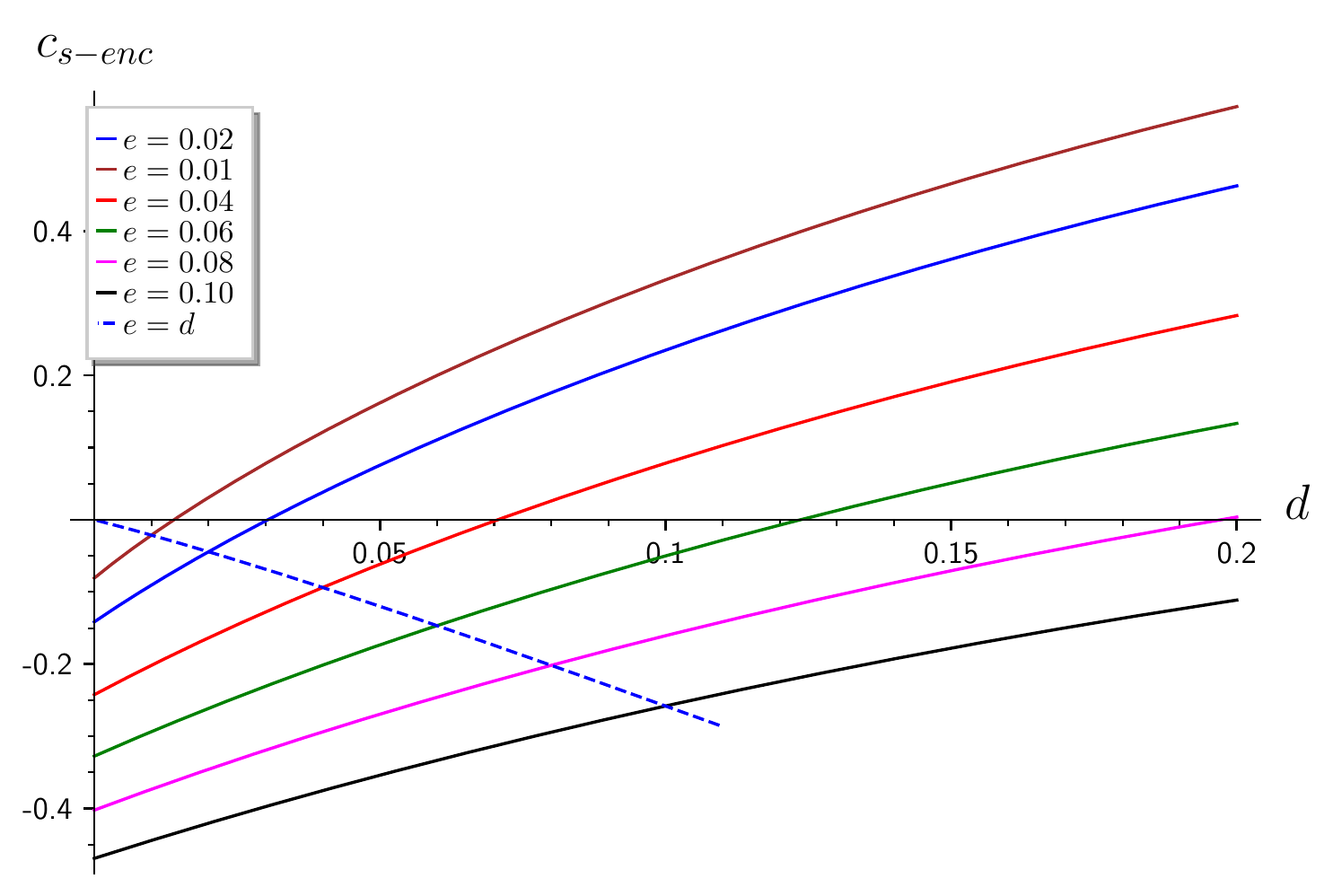}
  \caption{Secret key length balance (secret channel capacity) for reconciliation with encrypted $s_B$ enabling error correction offloading to untrusted parties.
Values are shown for different $e$. Left of the dashed line is the interesting region $d<e$, which has negative key balance and is therefore unfeasible.}
  \label{fig:keylength}
\end{figure}

We have seen, that to be left with a secure key after all steps, the outsourced error reconciliation has to hide either $\vec{s_e}$ or $\vec{s}_B$ with ITS properties.
However, $\vec{s_e}$ cannot be encrypted by masking, because the nature of the outsourced computation is to find a minimum weight vector which fulfills $\vec{e}\vec{H}^\top= \vec{s_e}$ for a given $\vec{s_e}$ and a public $\vec{H}$, 
which always requires to also publish a target vector $\vec{e}$ which is the reference for distance minimization.
Therefore, a simple solution is to encrypt $\vec{s}_B$ during transmission with previously acquired secure key material.
This requires $n H_b(e)$ additional key bits leading to a reduced capacity of $C_{s-enc} = H_b(e+d-2ed) - 2H_b(e)$.
Although the protocol is secure and enables offloading of error correction, it does not lead to positive key rate for the regions of interest where $d < e$, which is also shown in Figure~\ref{fig:keylength}.

\medskip

To give more evidence that an encrypted IR protocol with positive key balance is not achievable, we review most relevant and evident techniques to protect the key of Alice or even $\vec{s}_A$ in an ITS sense, to prevent Eve from learning Alice' key or increase $I(\vec{K}_A;\vec{K}_E)$.
In order to build an encrypted RR protocol, different techniques could be used, however, the parity check matrix $\vec{H}$ is considered to be publicly known, which limits the application of hiding techniques to the raw key vector.
Furthermore, the discussed solutions should not increase the computational effort to correct errors.

We start from the syndrome decoding equation $\vec{s_e} = \vec{s}_A - \vec{s}_B = \vec{e} \vec{H}^\top$ and discuss options to hide $\vec{s}_A$ from Eve, or to prevent from any increase in $I_{AE}$ by public discussion.
In order to hide the bit flip positions, we discuss the following additional techniques, which are evident approaches towards the security goals for offloading RR but also not providing any positive key rate.
\begin{itemize}
    \item Encrypting the raw key by one-time-pad (OTP), 
		\item permutation, and
    \item padding (i.e., adding dummy (error) bits).
\end{itemize}

\smallskip
\textit{Encryption.} 
If the goal is to hide $\vec{s}_A$ in an ITS way given $\vec{H}$ is public, either $\vec{s_e}$ or $\vec{s}_B$ must be one OTP encrypted.
$\vec{s}_B$ can be encrypted when transmitted to Bob or already at the key level, therefore ultimately hiding the key $\vec{s}_B' = (\vec{k} + \vec{m}) \vec{H}^\top$, where $\vec{m}$ is a random masking value, which must also be securely transmitted from Bob to Alice.
However, in the first case $|\vec{s}_B| = n H_b(e)$ bits are optimally required and in the second case number of raw key bits $|\vec{k}|$ are required, which is extremely inefficient.
Above, we have already shown that even the first case leads to negative key rates.

Unfortunately, encryption of $\vec{s_e}$ also cannot be used to hide bit error positions, because decoding requires a start vector to explore the vicinity to.
Finding a vector close to a random vector with public $\vec{H}$ leaks the bit flip positions $\vec{e}$ and therefore also $\vec{s_e}$.

\smallskip
\textit{Permutation.}
An alternative method to hide $\vec{e}$, $\vec{s_e}$ and thus $\vec{s}_A$ would be by permuting the raw key bits before running RR with an unencrypted $\vec{s}_B$.
Using a permuted key $\vec{k}' = \Pi(\vec{k})$ for the post-processing would render error correction information useless for Eve, however, has to be random for each block and applied on both peers in secret.
Thus, a huge amount of shared key material is required given the permutation has to be selected randomly from the $n!$ possible ones, which requires  $\mathcal{O}(n \log(n))$ bits to represent.
In the end, if the selected permutation has to be communicated over the public channel via OTP the key balance is even worse than with syndrome encryption.

\smallskip
\textit{Padding.}
Padding the raw key with dummy bits could be used to hide error bits if combined with permutation.
This corresponds to the technique of mixing raw key with dummy key bits.
However, also in this case the positions and value of the dummy key bits have to be agreed on secretly by Alice and Bob, which also requires too many bits.

Finally, additional errors could be introduced only at Alice.
Because the remaining error margin in practical CV-QKD is already very small, this technique can only hide small amount of information and substantially increases the computational work at the remote instance through the increased error rate.



In summary, all natural approaches for partially offloading RR with positive key rate to a single server in general seem unfeasible. 

\subsection{Verifiability of outsourced IR}\label{s:verifiability}

Besides the challenge of efficient yet secure outsourcing of information reconciliation, it is also important to have a means to efficiently check the correctness of the solution.
This prevents from actively malicious behavior of the remote instance doing the actual work.

Fortunately, the problem of error decoding comes with an efficient algorithm to check the result:
\begin{enumerate}
    \item Check if $\vec{e} \vec{H}^\top = \vec{s_e}$, abort otherwise
    \item (Optional) Check if the weight of $\vec{e}$ is indeed below the threshold of the code or is consistent with estimated error, abort otherwise.
\end{enumerate}

The firsts check can be easily computed by conducting the vector matrix multiplication and only requires additions modulo 2 (XOR) in the order of bits set in $\vec{H}$, which is especially efficient for LDPC codes.
The second check is even faster, if it can be performed for the used code.
The Hamming weight of $\vec{e}$ must be smaller than what can be corrected by the code.
However, not always can the correction capabilities of a code be bound, especially for often used LDPC this is not possible.
In such cases only the estimated error rate can be used to test the hypothesis of a bit flip vector being correct.
Nevertheless, there is still the final confirmation phase where an ultimate check is done to assure the key error probability, however, directly verifying the IR outsourcing step enables attribution of errors to external servers and flexible reaction besides aborting the whole process.

In summary, verifiability immediately follows from the nature of the problem.
This makes protection against malicious remote servers possible with minimal effort and does not require a full re-computation by Alice.

\subsection{Multiparty computation based outsourcing}

In the previous sections we presented an efficient solution for direct reconciliation (DR) offloading and discussed problems for RR in the single server model, i.e., with one remote server used for offloading.
Although relying on a single untrusted server seems the most desirable use case, it is natural to ask how efficient a multi-server configuration would be.
If multiple servers are available, secure multiparty computation protocols (MPC)---originally introduced by Yao~\cite{DBLP:conf/focs/Yao86}---can be used to obliviously compute arbitrary functions on sensitive data, thus they can also be used for CV-QKD.

MPC enables a set of parties to jointly evaluate a function, without leaking any information to any of the participating parties, beyond what can be derived from their own inputs and the computation result itself.
More precisely, MPC provides \emph{input secrecy} (or \emph{input privacy}), i.e., no party learns the input values of any of the other parties, and \emph{correctness}, i.e., the receiver of the result is ensured that the result is correct, even if some parties maliciously deviate from the protocol specification.
Furthermore, in an honest-majority setting with less than half corrupt servers, ITS secure protocols which achieve practical performance in many application scenarios exist.

We therefore looked into the problem of MPC-based information reconciliation based on generic MPC-based on secret sharing~\cite{DBLP:journals/cacm/Shamir79}.
If IR is done in MPC, the decoding can be done without learning anything about the error syndrome (private input) or vector (private output), but the parity check matrix can still be kept in clear.
In this model the peer offloading IR is encoding the error syndrome as private input for the MPC system, which computes the bit flip vector in a distributed form.
The different secret shares of the final result are then sent back to the peer, who can reconstruct it.

We study the practical efficiency for doing error decoding with low-density parity-check (LDPC) codes in an existing MPC framework to estimate the performance that can be achieved.
To the best of our knowledge, this is the first time this problem is considered: The only
known related work has been presented by Raeini and Nojoumian~\cite{Raeini2018}, who however only considered Berlekamp-Welch decoding for Reed-Solomon codes.

In general, we distinguish two main types of message-passing algorithms for LDPC decoding: bit-flipping algorithms and belief propagation~\cite{910577}.
The decoding approach typically used in QKD is from the category belief propagation (BP), and specifically uses sum-product mechanisms to update beliefs, an approach which works very efficiently but is not well suited for direct conversion to MPC.
This is because the algorithm works on floating point numbers and uses trigonometric functions in the belief update part.

Therefore, to initiate the research topic we focused on bit-flipping algorithms (BF) for our first approach.
BF algorithms have a very simple structure and work extremely fast, e.g., if implemented in hardware.
They are also well suited for MPC implementation and were selected for first benchmarking, although suffering from inferior performance in terms of information rate.

The bit-flipping algorithm is a non-probabilistic hard-input hard-output decoding algorithm
and works on the Tanner graph representation of the code.
The messages passing forth and back are all binary.
The main structure of the BF algorithm is similar for all variants.
In a first step, the variable nodes send their current value to the check nodes.
The check nodes feed back a bit to the adjacent variable nodes signaling if the check is valid.
After each variable node received the checkbits from all connected check nodes the current guess for the bit vector is updated.
Different approaches exist to update the variable nodes and to the best of our knowledge no optimized codes and methods for the particular case of QKD have been studied or analysed.
Therefore, we selected one to the most prominent solutions---Gallager's Algorihtm A and B---to demonstrate feasibility and applied it to existing codes used for BP algorithms.
The results of our first tests are shown in Figure~\ref{fig:mpcbitflip} which indicate that MPC-based real-time decoding for QKD is possible.

\begin{table}[ht]
  \centering
  \caption{Performance comparison of MPC version of bit flipping algorithm for LDPC decoding. The values show that the kilobit per second regime is feasible even without optimizations and block level parallelization.}
  \label{fig:mpcbitflip}
  \begin{tabular}{@{}l|c|c|c|c|c|c@{}}
  \toprule
  block size & bitwidth & circuit depth & time & data & rounds & bitrate@10iter \\
  & & & [s] & [MB] &  & [bps] \\ \midrule
  1000   & 4  & 9  & 0.06 & 3.0 & 65  & 1571 \\
  1000   & 8  & 11 & 0.09 & 3.8 & 80  & 1116 \\
  10000  & 4  & 9  & 0.11 & 4.4 & 85  & 8932 \\
  10000  & 8  & 11 & 0.14 & 4.7 & 95  & 7054 \\
  100000 & 4  & 9  & 1.2  & 44  & 805 & 8354 \\
  100000 & 8  & 11 & 1.4  & 47  & 846 & 7117 \\ \bottomrule
  \end{tabular}   
\end{table}

Clearly, to achieve the best performance, optimized codes must be studied and designed in tandem with MPC protocols \cite{LW2020}. 
Also, BP based alternatives to sum-product decoding should be studied to see how fast MPC versions of belief propagation methods can be pushed.
Additionally, for CV-QKD approximation approaches combined with multiedge-type codes \cite{PhysRevA.103.06241} seem promising for fast MPC implementation.
Nevertheless, our experiment already shows first results and paves the way for practical rates.

Additionally, optimization-based decoding would also be possible as an alternative to message passing algorithms, i.e., by leveraging linear programming (LP).
In LP decoding \cite{Feldman2005,Feldman2003DecodingEC} the maximum likelihood decoding problem is formulated as linear program.
Thus, it is possible to decode a symbol by solving an associated LP with conventional approaches, e.g. with a simplex algorithm where also MPC versions exist \cite{Toft2009}.
However, for the QKD use case with block sizes $k$ in the range of $10^4$ to $10^6$ bits and high error rates, the formulation would lead to a relatively large simplex tableau.
Very low rates can be expected for this solution approach given the measured performance for MPC-based LP solving reported in \cite{slotmachine-mpcopt}.

\section{Offload privacy amplification}\label{sec:privacyamp}

Privacy amplification (PA) is another important step in the post-processing stack, cf. Section~\ref{ssec:post-processing}.
It also requires a public channel for communication and is typically based on application of a randomly selected hash of a universal hash family, thus achieving information theoretical secure randomness extraction.
PA is used to extract the mutual information between Alice and Bob such that the adversary Eve is left without any information, except for a negligible error that can be made arbitrarily small.

Although the underlying matrix-vector multiplication seems rather efficient, because of finite key effects and its influence on the secure key rate large block length have to be used \cite{Leverrier2010}.
Therefore, also this step is computationally very demanding \cite{Wang2018a} and solutions to entirely offload this task from the device, or at least from the trusted area within a device, would be desirable.

In a PA protocol, Alice randomly selects a hash from a family of universal hashes with the right compression rate---based on Eve's potential knowledge on the key---and communicates the selected function publicly to Bob.
Both peers then apply the same function on the local reconciled key and arrive at the final shared key.
The main property of a universal hash family is that they guarantee a low number of collisions, even if the input is chosen by an adversary.

Because the block length in QKD is large, the complexity of the universal hashes is also relevant.
One family of strongly universal hashes is given by multiplication of the raw key with a random matrix which would need a lot of randomness.
Nevertheless, to reduce the randomness needed, a Toeplitz matrix can also be used for PA, which requires only $n+m$ random bits compared to the $n\cdot m$ for a random matrix.
The use of the Toeplitz matrix also reduces the computational effort for PA, because the diagonal structure also enables the use of a number theoretical transform for faster processing of the vector matrix product.

Assume that $\vec{k}'_A = \vec{k}'_B = \vec{k'}$ is the reconciled key at Alice and Bob, respectively, with length $n$ and $\vec{k}$ is the final keys of length $m$.
Then PA works as follows:
\begin{enumerate}
  \item Alice randomly generates a uniform string of length $n+m-1$ defining the Toeplitz matrix $\vec{T}$ and sends it to Bob.
  \item Alice computes $\vec{k} = \vec{k}' \, \vec{T}$ as final key.
  \item Bob receives $\vec{T}$ from Alice and also computes his key as $\vec{k} = \vec{k}' \, \vec{T}$.
\end{enumerate}
Thus, both parties do the same vector-matrix multiplication to shrink the identical keys from $n$ to $m$ bits, where the ratio $n/m$ for CV-QKD is computed as in Leverrier et al.~\cite{Leverrier2010} and for DV-QKD as shown by Scarani et al.~\cite{Scarani2009}.

If we want to offload PA, we would have to offload the core vector-matrix multiplication, which reduces the $n$ raw key bits to $m$ final bits.
The ratio is already known at the beginning of the PA step but the Toeplitz matrix has to be generated for each block and exchanged in clear, which makes offloading a problem.
Encrypting the PA matrix is not an option, it can be immediately seen that the key balance is negative if the encryption key for the matrix is longer than the raw key processed.
Thus, hiding the input/output keys while still offloading the computation is not feasible in a single sever-model.

However, if multiple servers are available, a very efficient non-interactive multi-party protocol is possible, i.e., without requiring the servers to communicate.
The protocol is shown in Figure~\ref{fig:rem-pa}.
The peer shares the raw key into $n$ parts with a linear secret sharing scheme---working over $\mathbb{F}_2$ or a larger prime field $\mathbb{F}_p$---, and sends them to the servers (one share per server).
The servers compute the $[\vec{k'}] = [\vec{k}] \vec{T}$ where $[.]$ denotes the sharing of a value.
Because of the linearity of the secret sharing scheme, the necessary multiplications and additions can be done on the shares without interaction between the servers.
The results are then sent back to Alice who reconstructs the final key.
For the case of prime fields, Alice additionally reduces the result vector ${mod}\ 2$.

\begin{figure}[t]
  \textbf{Protocol \texttt{REM-PA}:}\\[0.5em]
  QKD-Peer $P_Q$:\\[-1.3em]
    \begin{enumerate}
      \item Generates a sharing $[\vec{k}]$ of key $\vec{k}$ by calling \textit{share} of a linear secret sharing algorithm outputting $n$ secret shares $[\vec{k}]_i$ for $i = 1, ..., n$ 
      \item Sends shares $[\vec{k}]_i$ to Party $P_i$
      \item Receive enough shares $[\vec{k}']_i$ to \textit{reconstruct} $\vec{k}'$
      \item Reduce elements of vector $\vec{k}' {mod}\ 2$ for prime fields
      \item Return $\vec{k}'$
    \end{enumerate}
    MPC-Party $P_i$:\\[-1.3em]
    \begin{enumerate}
      \item Receive share of key string $[\vec{k}]_i$
      \item Calculate $[\vec{k}']_i = [\vec{k}]_i \vec{T}$ with conventional PA algorithm
      \item Send $[\vec{k}']_i$ back to Q
    \end{enumerate}
    \caption{\texttt{REM-PA} Protocol}
    \label{fig:rem-pa}
\end{figure}

The security of the protocol against passive adversary is governed by the security of the underlying secret sharing scheme:
Because the parties do not interact with each other but only communicate with the peer, they cannot learn any information about the final key as long as an ITS linear secret sharing is used, e.g., additive or Shamir secret sharing~\cite{DBLP:journals/cacm/Shamir79}.
The computational effort of the solution is the same for every server, which would also be the same as for local computation.

Unfortunately, the \texttt{REM-PA} protocol does not provide efficient verifiability beside re-computation or spot checking, and therefore efficient protection against active attackers cannot be easily achieved during the PA phase.
However, if the confirmation round is shifted after PA it will detect errors in the keys and prevent from erroneous keys by aborting the protocol.
Thus, it can also detect malicious behavior of the external servers, but not directly attribute the errors to them.
If the confirmation is shifted after PA it is important to encrypt the tag sent because otherwise the leaked information cannot be removed anymore as is normally done by PA.
Additionally, a secure channel is assumed to distribute the shares to the severs, which may prevent from certain use cases.
However, contrary to the MPC-based LDPC decoding, no interaction between servers is required.


\section{Use Cases}\label{sec:usecases}

To answer why offloading computationally intensive tasks is interesting at all, we present the expected benefits in general and discuss advantages for certain networking scenarios.

The overall goal which can be achieved are savings in energy and/or cost at the device side
beneficially impacting the cost-effectiveness of the end-user equipment.
Therefore, if the devices are simpler and require less computational power, the cost savings could be substantial, e.g., in cases where the user buys the equipment.
Compared to data center environments, the devices are also less energy-efficient for running computation-intensive tasks.
If this part can be offloaded to a more efficient data center also the overall operational cost can be lowered in addition.
Therefore, dedicated cloud solutions which further pool information reconciliation for a larger amount could further help to reduce energy consumption.
By regularly updating the external hardware resources, the system can benefit from Moore's law and the continuous drop in cost of compute resources.
They can even be shifted flexibly between different locations and data centers to optimize energy usage and cost if more offerings are available.
In general, it would be even possible to leverage public cloud services for REM-IR, which requires no trust assumption at all about the environment.
Because of these arguments, we think the ability to offload and relocate computationally intensive tasks also leads to higher energy efficiency of compute resources.

Additionally, it could also lead to more flexibility on the QKD level, i.e., QKD as a service.
The virtualization of the computationally expensive post-processing tasks could be convenient in the future.
Not all optical network units (ONU) have access to QKD functionality but the same hardware may be used for coherent passive optical networks (PON) and CV-QKD~\cite{DBLP:conf/ecoc/MilovancevHVLHS21,phlox}.

So, we may provide QKD to them by just allocating additional processing resources while also switching their software-defined transceiver into QKD mode.
Furthermore, networking equipment is installed for longer times and not often updated.
This is even more true for high-cost security-certified equipment, because upgrading security-certified equipment is a cumbersome and costly process which typically requires re-certification.
Being able to update certain non-critical components without needing to exchange or re-certify the core QKD device hardware can greatly simplify the upgrade process.


To show how the advantages relate to concrete use cases we quickly mention three examples.

\smallskip
\textit{Access networks.}
In the case of access networks, we find constraint resources (computing energy) and the network units must be low-cost because they are the driving cost factor, especially, because only very low key rates are typically required (AES key refreshing).
If computational resources are pooled in such a scenario, cost can be substantially reduced, not just in case of a reduced user subscription ratio, but also due to time sharing of centralized CPU resources.
Furthermore, because the optical part is low energy the dominant cost and energy factor is when a CPU is partially idle, which should be avoided.

\smallskip
\textit{Satellite communication.}
Satellites have a particularly long lifetime (20-30 years) and have to be remotely operated and maintained.
They also have limited access to energy resources and reducing energy consumption is of paramount interest.
This is especially true if low-cost (mobile) earth stations should be supported or even for inter satellite links.
Offloading post-processing can make satellite transceiver possible and increase the connectivity for individual satellites.

\smallskip
\textit{Integrated COTS Hardware.}
Finally, beside the evident advantages of outsourcing protocols like REM-IR to data centers, the concept can also be interesting when applied within the device.
QKD devices are complex systems \cite{TPH+2009b} and comprise many different components which makes security auditing and certification very hard.
To achieve strong security guarantees only trustworthy hardware and software can be used to process key material in plaintext \cite{LQM+2008}.
Furthermore, to prevent from side channel attacks and backdoors it would be desirable to reduce the amount of trusted components and the complexity of the secure environment in a device as good as possible.
Therefore, if the components processing sensitive key material can be reduced, this results in a smaller attack surface, simplifies security analysis and helps in the certification process.
The MPC-based protocols presented can be used for this purpose, i.e., to reduce the trusted environment on the device architecture level with all its benefits.
Within a device it is also feasible to realize the secure channels required in the MPC model.
Thus, it would allow for the integration of COTS hardware in QKD systems only processing keys in encrypted form.

\section{Conclusions}\label{sec:conclusions}

In this work we introduced the idea of offloading information reconciliation and privacy amplification steps of QKD post-processing.
These are the two computationally intensive tasks in processing raw key measurements to secure shared key between two QKD peers.
We show that outsourcing information reconciliation is possible and straightforward for DV-QKD even in a single server model and against an active adversary.
However, for CV-QKD, which leverages reverse reconciliation to overcome the 3dB transmission bound, the same is not true.
We also give an intuition that it is not possible in general to achieve positive key rates with a single server and analyze potential performance in a multi-server setting. 
We also look into privacy amplification, where we propose a protocol for multiple servers.
Finally, we motivate potential benefits and discuss use cases where this approach is relevant.

Proving the impossibility of single-server PA offloading as well as \textit{weak offloading} is left for future work.
Additionally, MPC optimized versions of sum-product decoders are currently under investigation and will be presented in follow-up work.

\section{Patents}

The basic scheme \textit{IC-REM} from this work was first patented in Austria (AT519476B1) and later also in Europe (EP3607446B1) and US (US11128445B2).
However, only in this work we provide the security analysis and additional methods as well as the limitations for the technology.

\vspace{6pt} 

\authorcontributions{Conceptualization, T.L. and B.S.; methodology, T.L, S.K., C.P. and B.S.; validation, T.L., S.K. and C.P.; formal analysis, T.L. and S.K.; investigation, T.L.; writing---original draft preparation, T.L., S.K., and C.P.; writing---review and editing, T.L., S.K, C.P. and B.S. All authors have read and agreed to the published version of the manuscript.}

\funding{This research has received funding from the European Union’s Horizon 2020 research and innovation programme under grant agreements No 857156 (OPENQKD) and No 830929 (CyberSec4Europe).}

\conflictsofinterest{The authors declare no conflict of interest.}

%

\begin{adjustwidth}{-\extralength}{0cm}

\reftitle{References}


\bibliography{bib/refs}

%


\end{adjustwidth}
\end{document}